\documentclass[a4paper,fleqn]{cas-dc}

\usepackage[numbers]{natbib}
\def\tsc#1{\csdef{#1}{\textsc{\lowercase{#1}}\xspace}}
\tsc{WGM}
\tsc{QE}
\tsc{EP}
\tsc{PMS}
\tsc{BEC}
\tsc{DE}

\begin{document}
\let\WriteBookmarks\relax
\def\floatpagepagefraction{1}
\def\textpagefraction{.001}
\shorttitle{Generating Ultrasonic B-scans with Generative Adversarial Network}
\shortauthors{Posilović et~al.}

\title [mode = title]{Generative adversarial network with object detector discriminator for enhanced defect detection on ultrasonic B-scans}

\tnotetext[1]{Corresponding author at: University of Zagreb, Faculty of Electrical Engineering and Computing, Croatia}

\author[1]{Luka Posilović}[orcid=0000-0003-2639-0812]
\cormark[1]
\ead{luka.posilovic@fer.hr}
\credit{Conceptualization, Methodology, Software, Validation, Data Curation, Writing - Original Draft}

\author[1]{Duje Medak}[orcid=0000-0001-6261-206X]
\ead{duje.medak@fer.hr}
\credit{Software, Data Curation, Writing - Review and Editing}

\author[1]{Marko Subašić}[orcid=0000-0002-4321-4557]
\ead{marko.subasic@fer.hr}
\credit{Conceptualization, Resources, Writing - Review and Editing, Supervision}

\author[2]{Marko Budimir}[orcid=0000-0001-6508-1305]
\ead{marko.budimir@inetec.hr}
\credit{Resources, Data Curation, Writing - Review and Editing, Funding acquisition}

\author[1]{Sven Lončarić}[orcid=0000-0002-4857-5351]
\ead{sven.loncaric@fer.hr}
\credit{Resources, Writing - Review and Editing, Supervision, Project administration, Funding acquisition}

\address[1]{University of Zagreb, Faculty of Electrical Engineering and Computing, Zagreb, Croatia}
\address[2]{Institute for Nuclear Technologies (INETEC), Zagreb, Croatia}

\begin{abstract}
Non-destructive testing is a set of techniques for defect detection in materials. While the set of imaging techniques are manifold, ultrasonic imaging is the one used the most. The analysis is mainly performed by human inspectors manually analyzing recorded images. The low number of defects in real ultrasonic inspections and legal issues considering data from such inspections make it difficult to obtain proper results from automatic ultrasonic image (B-scan) analysis. In this paper, we present a novel deep learning Generative Adversarial Network model for generating ultrasonic B-scans with defects in distinct locations. Furthermore, we show that generated B-scans can be used for synthetic data augmentation, and can improve the performance of deep convolutional neural object detection networks. Our novel method is demonstrated on a dataset of almost 4000 B-scans with more than 6000 annotated defects. Defect detection performance when training on real data yielded average precision of 71\%. By training only on generated data the results increased to 72.1\%, and by mixing generated and real data we achieve 75.7\% average precision. We believe that synthetic data generation can generalize to other challenges with limited datasets and could be used for training human personnel.
\end{abstract}

\begin{graphicalabstract}
\includegraphics{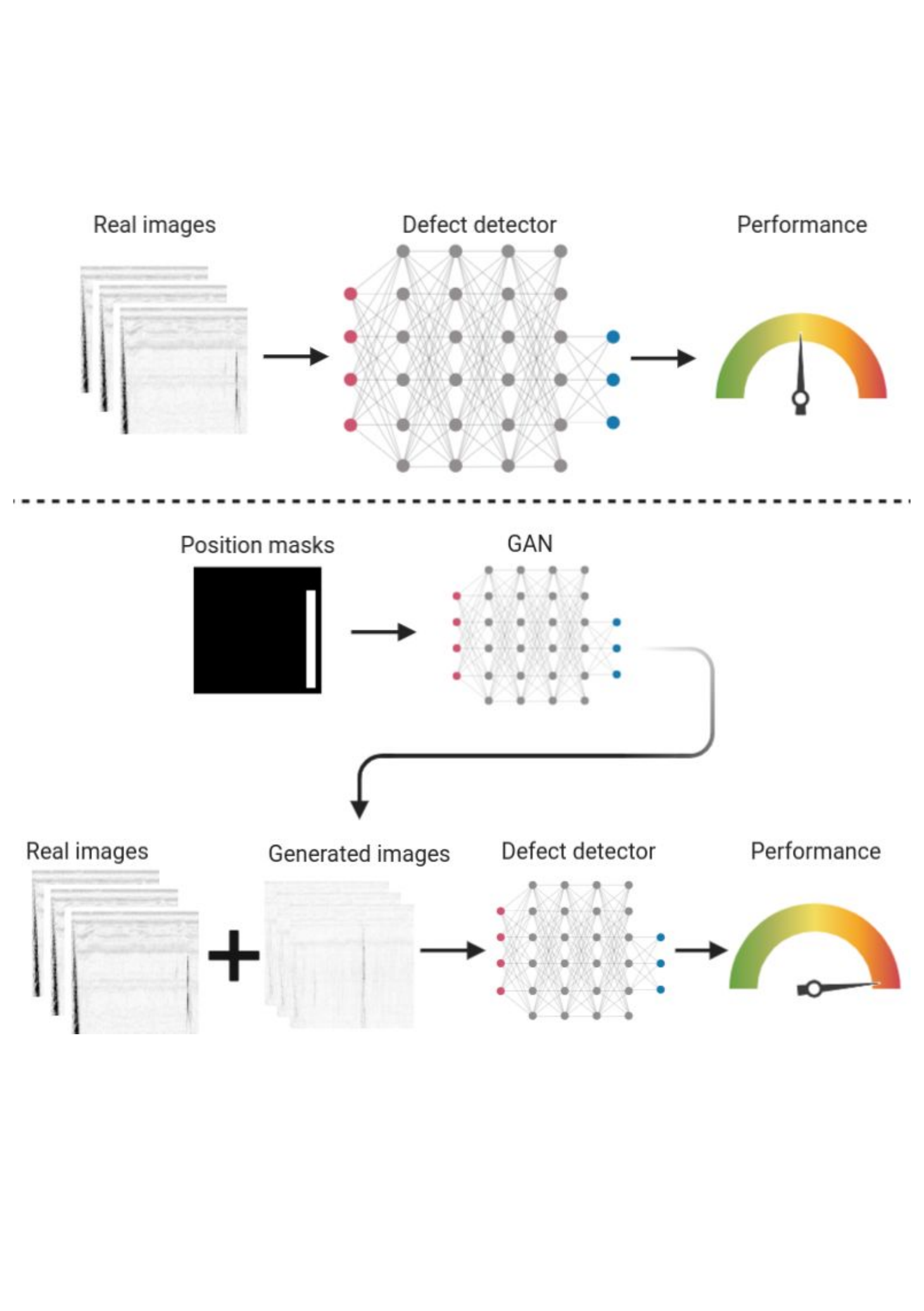}
\end{graphicalabstract}

\begin{highlights}
\item A novel GAN architecture for generating high quality images with objects at precise locations
\item Our proposed GAN is able to generate highly realistic data that can help improve the object detector
\item An improvement of almost 5\% of average precision was achieved when training on a combined dataset of real and images generated with our GAN
\end{highlights}


\maketitle

\section{Introduction} \label{introduction}

Non-destructive testing (NDT) is widely used in science and industry to evaluate properties of materials, components or systems without causing damage \citep{cartz1995}. Many different methods are available such as visual examination, ultrasonic, eddy current, to name a few. Among them, ultrasonic testing (UT) stands out due to its versatility. Highly sensitive on most materials \cite{Ye_2018}, high signal to noise ratio \cite{broberg2014imaging}, and ability to determine defect location and type \cite{Ye_2018} are some of its advantages. Ultrasonic data can be represented in several different formats suitable for analysis including A, B, or C-scans \cite{krautkramer1983ultrasonic}. An A-scan is an ultrasonic signal's amplitude put as a function of time, B-scan displays a cross-sectional view of the inspected material, and a C-scan provides a top view of its projected features \cite{posilovic2019flaw}. During analysis, inspectors use all of them in order to make a decision and evaluate the data.

Automated analysis has long been used in many NDT systems. However, so far it has been limited to classical decision-making algorithms such as amplitude threshold \cite{virkkunen2019augmented}. Complex data such as the one from ultrasonic inspection makes it hard to develop an automated analysis. All ultrasonic analysis is, to the best of our knowledge, done manually by a trained human inspector. It makes ultrasonic analysis highly reliant on the inspector’s experience. The automated analysis could make the process much faster and reliable. There have been some attempts in developing an automated UT analysis \cite{Souza2019, Munir2018, meng2017ultrasonic, virkkunen2019augmented, posilovic2019flaw}, but very few of them involve using deep learning and modern deep convolutional neural networks (CNNs) on B-scans. The prerequisite for using deep learning is a large, annotated dataset. Due to the low number of flaws in real ultrasonic inspections and legal issues considering data from such inspections available data is limited. Data is the biggest drawback in the development of proper automated/assisted ultrasonic analysis. This challenge can also be found in many medical image analysis tasks \cite{shorten2019survey} where, due to the rarity of some pathology and patient privacy issues, data availability is very modest. Furthermore, unlike medical datasets, there are no publicly available UT datasets. 

Researchers attempt to overcome this problem by using transfer learning \cite{ventura2007theoretical} in combination with freezing the backend CNN layers \cite{soekhoe2016impact} which is shown to enhance the accuracy of models. Using data augmentation is also the standard procedure for network training. However, data augmentation methods are limited and only slightly change some aspects of existing images (e.g. brightness modulation). Very limited additional information can be gained by such modifications. Synthetic data generation of high-quality images is a new type of state-of-the-art data augmentation \cite{frid2018gan}. Generative models such as generative adversarial networks (GANs) offer more variability and enrich the dataset to further improve the training process. 

In this work, we have presented a novel GAN architecture for generating high quality and realistic UT B-scans. Afterward, we have used the generated images to train an object detection neural network to detect defects in real images. We have compared training the object detector with only real images using traditional data augmentation and with GAN generated images. We have also developed a more traditional method of generating new images using the Copy/Pasting technique which proves the effectiveness of the GAN.

\subsection{Contributions}

The main contributions of this work are the following:
\begin{itemize}
    \item a novel GAN architecture for generating high-quality ultrasonic images with objects at precise locations,
    \item experimental demonstration that expanding the ultrasonic dataset with generated synthetic data increases the performance of the defect detector.
\end{itemize}

\subsection{Related work}

Data availability is a major problem when using deep learning for defect detection. B-scans are the ideal data representation for accurately detecting defects and further estimating their depth and size. However, a B-scan usually consists of hundreds of A-scans which further aggravates the problem of lack of data. Developed algorithms for defect detection can be divided into three groups related to data representation being used; A-scans \cite{BETTAYEB2004, Sambath2011, Chen2014, AlAtaby2010, Matz2006, KHELIL2005, MENG2017, Souza2019, CRUZ2017, guarneri2013, Munir2018, Veiga2005}, B-scans \cite{cygan2003, kechida2012, virkkunen2019augmented, posilovic2019flaw} and C-scans \cite{kieckhoefer2008image, dogandzic2007bayesian}. The A-scan analysis is the most researched group of all which is also related to the data problem. Developed algorithms mostly include a combination of wavelet transform \cite{BETTAYEB2004, Sambath2011, Chen2014, AlAtaby2010, Matz2006, KHELIL2005, meng2017ultrasonic}, discrete Fourier transform \cite{Souza2019, CRUZ2017} or discrete cosine transform \cite{CRUZ2017} and a support vector machine or artificial neural network classifier. B-scans keep the geometrical coherence of the defect which leads to a better noise immunity \cite{cygan2003}. However, the analysis of B-scans can only be seen in a few works \cite{posilovic2019flaw, virkkunen2019augmented}.  In \cite{posilovic2019flaw} two popular deep learning object detection models, YOLOv3 and SSD, have been used for defect detection. In \cite{virkkunen2019augmented} a deep learning detector has been tested on augmented images, but with only three defects in the specimen block. Regarding C-scans, in \cite{kieckhoefer2008image} a method based on the comparison of the scan with a reconstructed reference image has been made. The method was able to detect all defects in their dataset, but with a high number of false-positive detections. There have also been some attempts in estimating defects from noisy measurements using Bayesian analysis \cite{dogandzic2007bayesian}.

There have been some attempts in using data augmentation to enlarge the existing dataset. As mentioned, in \cite{virkkunen2019augmented}, although only three defects were present in the test block, a copy/pasting data augmentation has been used to enlarge the dataset for training a deep learning detector. There are many variations on pasting and blending objects on the background in order to make the images look as realistic as possible. For instance, it can be done using Gaussian blur or Poisson blending \cite{perez2003poisson} to smooth the edges. In \cite{dwibedi2017cut} a comparison between different merging techniques has been made, using a combination of blending methods performed the best for most objects. On the other hand in \cite{rao2017cut} authors have pasted objects on random backgrounds and achieved improvements without any blending. Finally, generative adversarial networks (GANs) have recently become a popular topic of research in the field of synthetic data generation and augmentation. GANs were first conceptualized in \cite{goodfellow2014generative} in 2014. They can be used to generate images, video, audio, text, and much more. The development of the GAN came a long way in a short period of time. There are many different GAN architectures. An interesting approach to GANs are image to image translation models. They are used for style transfer between images \cite{zhu2017unpaired}, image inpainting \cite{liu2018image} and even generating images from masks \cite{isola2017image}. One of the examples of those models is the Pix2pixGAN \cite{isola2017image} and its successor pix2pixHD \cite{wang2018high}. GANs show promising results in generating realistic images for human faces from noise with StyleGan2 \cite{Karras2019stylegan2} or converting position mask images to street-view with Pix2pixHD. A lot of work has been done for enlarging data sets in medical imagery. Pix2pixHD has proved to be useful in generating skin lesion images using semantic label maps \cite{bissoto2018skin}. An Inception-v4 classifier \cite{szegedy2016inception} has been trained using real and combined real and data generated with the Pix2pixHD. Training the classifier on a combined real
and generated data achieved a 1\% improvement of the area under the ROC curve. In \cite{frid2018gan} authors have applied the GAN framework to synthesize high-quality liver lesion images for improved classification. In \cite{bi2017synthesis} authors have developed a multi-channel GAN (M-GAN) to generate PET images from CT scans. A similar approach with a cGAN has been made in \cite{ben2017virtual, ben2019cross}. With generated data, they have achieved a 28\% reduction in average false positive per case. Generating MR images from CT scans with paired and unpaired data has been researched in \cite{jin2019deep}. An MR-GAN with a concept similar to CycleGAN \cite{zhu2017unpaired} has been developed for the purpose. In \cite{kazuhiro2018generative} a DCGAN has been employed to generate realistic brain MR images. Data augmentation using non-convolutional GAN has been tested on three different non-image datasets \cite{tanaka2019data}. Generated data has performed even better than real data when classifying using a Decision Tree (DT) classifier.

\subsection{Outline}
This paper is organized as follows. Section \ref{dataset} gives a detailed description of the dataset. Section \ref{experiment} demonstrates the experimental procedure. Proposed GAN architecture and copy/pasting method are presented in Section \ref{methods}. Results are shown in Section \ref{results} and a conclusion is given in Section \ref{conclusion}.

\section{Dataset} \label{dataset}

\begin{figure}[t]
    \centering
    \includegraphics[width=0.48\textwidth]{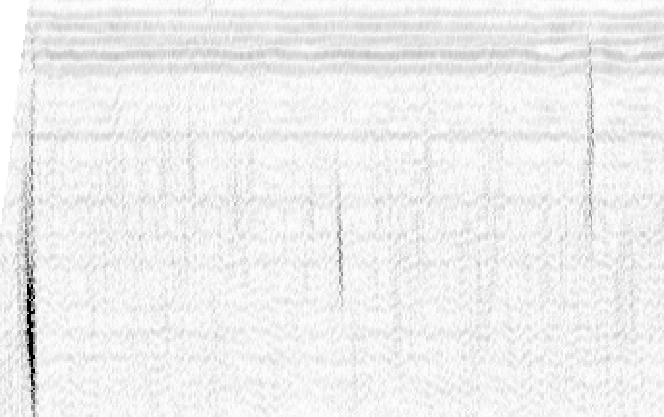}
    \caption{Example of an ultrasonic B-scan with defects}
    \label{fig:Bscan}
\end{figure}

The dataset was obtained by scanning six steel blocks containing artificially created defects in the internal structure. Blocks varied in size and contained between six to 34 defects. In total there were 68 defects. Blocks were scanned using INETEC Dolphin scanner with phased array probes. An INETEC phased array ultrasound transducer of central frequency of 2.25MHz was used. Data were acquired by scanning from an angle of 45 degrees to 79 degrees with a 2-degree increment. Blocks were also scanned with a skew of zero and 180 degrees. INETEC SignyOne data acquisition and analysis software was used to process the data and create B-scans (further noted as images) that were used in the dataset. Data were converted to B-scans as-is, without pseudo-coloring, as grayscale images. All images were then converted into patches of size 256x256 pixels and annotated by multiple human experts. There were in total 3825 images with a total of 6238 annotations. Dataset was split into train, test, and validation subsets. The train consisted of 2278 images, validation of 379 images, and the test of 1168 images. Details of the dataset split can be seen in Table ~\ref{table:dataSplits}. The split was made such that defects in each subset are unique, they do not appear in other subsets. GAN and the object detector were trained using this train/validation/test dataset split. As for the copy/pasting method, we copied the defects from the training subset in the form of the rectangle patches from the annotations and pasted them on the empty UT backgrounds. There were 3400 empty UT backgrounds from all of the blocks and 4283 defect patches from the training subset.

\begin{table}[width=.9\linewidth,cols=6,pos=ht]
\caption{Number of images and annotations in each of the data subset}\label{table:dataSplits}
\begin{tabular*}{\tblwidth}{@{} LLLL@{} }
\toprule
{} &
TRAIN &  
VALIDATION & 
TEST\\
\midrule
\midrule
Number of images & 2278 & 379 & 1168 \\
Number of annotations & 4283 & 745 & 1210 \\
\bottomrule
\end{tabular*}
\end{table}

\section{Synthetic data generation} \label{experiment}

The acquired dataset might not be enough to properly train an object detector to detect defects. For this reason, we propose two methods to expand our dataset with synthetic data.

In this section, we have described the procedure of the experiment in this work. We developed two methods for synthetic image generation and use a well-established object detector to test the quality of the generated data. We first start by describing the more traditional method for generating images and proceed to describe a deep learning approach with our GAN. We then explain the usage of the object detector in the experiment.

\begin{figure}
    \centering
    \includegraphics[width=0.48\textwidth]{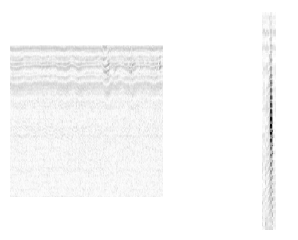}
    \caption{An example of an empty canvas and a randomly extracted defect}
    \label{fig:CP_input}
\end{figure}

Our first generative method is a copy/pasting (C/P) technique. Copy/pasting is a very logical method for enlarging the ultrasonic dataset because of the large number of B-scans without defects. We call these images canvases because we paste extracted defects on them. We extracted all of the defects from the training set and pasted them on canvases in random locations. The exact method is explained in the next section. An example of an empty image canvas and an extracted defect can be seen in Figure \ref{fig:CP_input}. 

\begin{figure}
    \centering
    \includegraphics[width=0.48\textwidth]{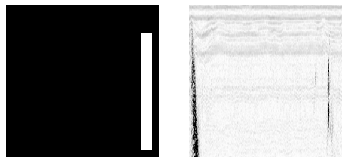}
    \caption{A pair of GAN input image and a desired output (real image)}
    \label{fig:GAN_input}
\end{figure}

The second method we propose is our own GAN architecture for the purpose of generating UT B-scans. Our GAN is an image-to-image GAN, with paired position masks as input and real images as a desired output. We make position masks from all annotated images in the training set. An example of a paired position mask and the real image is shown in Figure \ref{fig:GAN_input}. Position masks on the input of the generator serve as a location label for the desired position of the defect on the generated image. The main novelty of our GAN is the usage of a pre-trained object detector for training the GAN. We use the object detector as an additional discriminator to provide information on the quality of the defect on the generated image when compared to the real image. It is important that the defect is positioned accurately as drawn in the position mask and that it is merged well with the background. After training the GAN we generated new position masks used for generating synthetic data. For generating position masks we extract the aspect ratios of all annotations from the training set. We then create random masks by randomly putting the white boxes with predefined aspect ratios on black canvases. Our generated images have between one and four defects per image.

For estimating the quality of the generated images we used a popular object detector that was already proven in UT defect detection on images \cite{posilovic2019flaw}, a YOLOv3 \cite{yolov3} object detector. We first trained the detector using only real images and some traditional augmentations explained in the next section. We then tried training the object detector with images generated using the copy/paste method. We also trained the detector with a combination of real and generated images. We used the object detector trained on real images for training the GAN. We generated synthetic data with our GAN and again trained the object detector with generated images and a combination of real and generated data. Each of the trained versions of the object detector was tested on the same test dataset described in the previous section. Also, the same validation set was used in all three training variations.

\section{Methods} \label{methods}

In this section, a detailed explanation of developed methods is given. First, the copy/pasting method is described. Then the architecture of our proposed GAN is described with all of its special features. In the end, a short overview of the used object detector is given.

\subsection{Copy/paste method}

This method is a simple method that serves to illustrate the complexity of generating synthetic data. While these images might look visually appealing, they are not of the same quality as the ones generated by the GAN.

As mentioned in the Section \ref{dataset} we have previously extracted defects from images in the training set. We paste them on random locations on images without visible defects. The process goes as follows. First, we randomly pick a canvas and randomly select the defect we would like to paste on it. We then put a threshold on a defect image. We make a binary pseudo mask by creating a binary image from the thresholded image and dilate it for two iterations. We then use the mask to extract only the defect from the initial defect patch image. We randomly select the position where we will paste the defect and calculate the compatibility of the selected defect background and the canvas on that location. We calculate the compatibility by calculating an average value of the background of the canvas and the defect. If these two values do not differ by more than 5\%, we accept the proposed location. If these two values differ by more than 5\% we try to select another location. We then select another image/canvas pair and repeat the process. For each new image, we set the limit of 100 attempts after which we just move on to generate another image. Usually, this limit is enough to never move on, but always be able to find the right canvas/defect and location pair. If we found the right pair, we then proceed to paste the defect on the canvas. We first adapt the brightness of the defect to even further match the one from the canvas. We calculate the brightness of the location on the canvas and adapt the brightness of the defect to it. All that is left after that is to paste the defect. We concatenate the canvas and the defect, calculate the per pixel minimum of the two and merge it into one resulting image.

\begin{figure*}
    \centering
    \includegraphics[width=\textwidth]{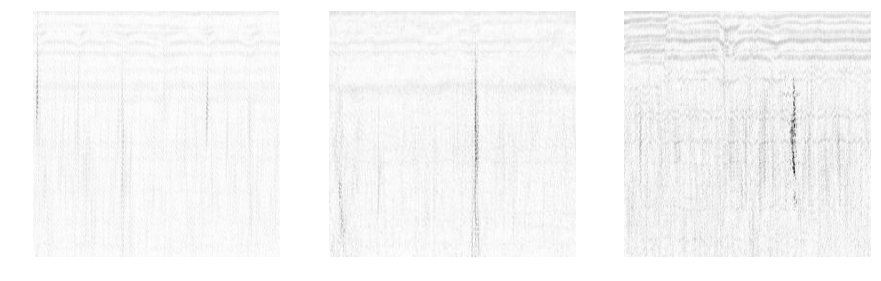}
    \caption{Samples of (from left to right): real image, image generated with GAN and a copy/paste image}
    \label{fig:real_gan_cp}
\end{figure*}

Samples of the real image, image generated with our GAN, and image generated with a copy/paste method are shown in Figure \ref{fig:real_gan_cp}. 

\subsection{GAN}

\begin{figure*}
    \centering
    \includegraphics[width=\textwidth]{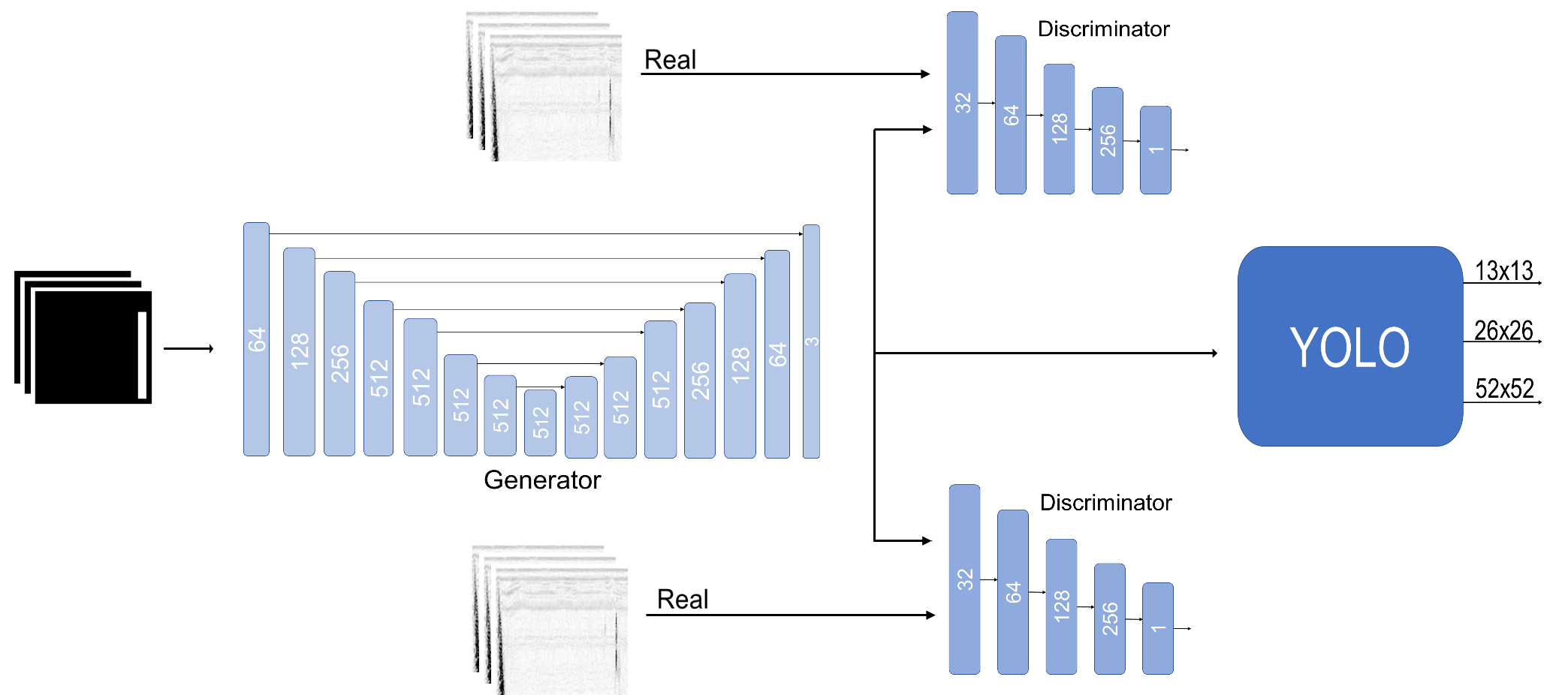}
    \caption{Simplified architecture of our DetectionGAN}
    \label{fig:DetGAN}
\end{figure*}

 The basic idea of GAN is a combination of two neural networks, a generator that generates high-quality images from random noise and a discriminator which tries to distinguish generated images from the real ones. The constant rivalry between the generator and the discriminator is what makes GANs adversarial. Mathematically, discriminator and generator play a minimax game where the goal of the generator is to maximize the probability of discriminator labeling generated images as real samples and discriminator has the goal of minimizing that probability while being able to label real data as such. This neural network configuration enables unsupervised learning of both generator and discriminator.

We call our GAN the DetectionGAN for its specific architecture. We base it on the Pix2pixHD implementing some of the features from it. Our GAN consists of a U-net generator, two PatchGAN discriminators \cite{isola2017image} that work on different scales and a pre-trained object detector which serves as an additional discriminator. We train the GAN with image pairs of real images and their position masks. 

Input to the generator is the position mask defining the position of the defect. The output of the generator is connected to the discriminator and to the object detector. There is a total of 54,409,603 parameters in the generator. All of them are randomly initialized and trained. Unlike in Pix2pixHD, we do not have a two-stage generator nor we upscale the position mask in order to generate a higher resolution image. 

Discriminator has a position mask image concatenated to the generated or real image as an input. Image and mask are concatenated across channels axis. The goal of the concatenation of the position mask and the image is to provide information on the position of defects in the image for the discriminator. This concatenation provides an improvement shown in the Section \ref{results}. Discriminator gets the real and the generated image during each step as an input. Since there are two discriminators, one of them inputs images downscaled by a factor of two in order for them to work on different scales. They have 1,391,554 parameters that are randomly initialized and trained. 

For the additional discriminator we use a YOLO object detector during this experiment, but any other object detector could be used. It helps generate highly realistic images with defects in precise, desired locations. We input the generated image and then the real image pair and compare the outputs. We want these two outputs to be the same so that there is no difference between the generated and real image for the detector. This way we ensure defects are placed on the exact locations and without any artifacts. Using an object detector as a discriminator provides a significant improvement shown in the Section \ref{results}.

The simplified architecture of our GAN is pictured in Figure \ref{fig:DetGAN}. Filter sizes at each layer of the generator and discriminators are noted in the figure.

\subsection{Object detector}

Our object detector, You Only Look Once (YOLO) version 3 is mainly taken from \cite{posilovic2019flaw} where it proved to be able to detect defects with high average precision. With this work, we used the proposed object detector to improve its performance on a more realistic dataset. We train the object detector on real, generated data and a combination of those two. We train the networks with the same hyperparameters in order to have a fair comparison. We first trained the detector on real data, tuned the hyperparameters to achieve the best possible performance and used the proposed parameters for training with other data combinations. We input images of size 416x416 pixels. We used a pre-trained backbone and froze its parameters while training. Hence, we trained only 20,974,518 of a total number of 61,576,342 parameters. 

\section{Experimental setup and results} \label{results}

\subsection{Experimental setup}

In this section, we describe the experiment and parameters used to train our GAN and the object detector. Our experiment goes as follows. We first train an object detector with real data. We then used that neural network to train our GAN. We generated synthetic images using the copy/pasting method and the deep learning GAN method. We generated 200,000 synthetic images with both copy/paste and GAN method. For each version of GAN we use the same position masks to generate synthetic images. We again train the object detector using the generated data and compare results.

Position masks for the input of the GAN are of size 256x256 pixels, as well as the generated images. For training the generator we use Adam optimizer with a first moment term of 0.5, the second one of 0.999, and a learning rate of 0.0002. One of our discriminators has an input image of 256x256 pixels, while the other one has a downscaled image of 128x128 pixels as an input. We train discriminators using Adam optimizer with the same parameters as the generator. We train our GAN using a set of loss functions. For the generator, we use four different losses. At the output of the generator, we calculate the L1 loss on the generated image and the paired real image. For propagating discriminator output to the generator we use the mean squared error loss. We also use the feature map L1 loss in the generator similar to the one in \cite{wang2018high}. We also compare the output of the object detector when inputting the real image and the output from the generator. We use the L1 loss to propagate it down to the generator. For training the discriminator we use the mean square error loss, just like it is used in Pix2pixHD. We use the pre-trained object detector and do not train it during training the GAN. We implement a set of simple data augmentations for training the GAN including horizontal flipping, brightness modulation, and random cropping. These data augmentations enable us to achieve great results in training the GAN and generating high-quality images. We train the GAN for 800 epochs and for the last 100 epochs we linearly reduce the learning rate to zero. Each epoch corresponds to one sweep through all the images in the training dataset. We trained with a batch size eight. It took us nearly 96 hours to train the GAN using a single NVIDIA RTX 2080Ti graphics card.

For training the object detector we used the following configuration. We use batch size eight and Adam optimizer with a learning rate 0.003. Custom anchors were calculated on the training set for all of the training combinations. We slightly changed only the ignore threshold hyperparameter to 0.6 from the original YOLO implementation. We used checkpoints while training the model. An early stopping callback was used to stop the training after the validation loss didn't improve for over eight epochs. We reduced the learning rate after every two epochs with no improvement on the validation set. We also used some basic augmentations while training all of the models. Those augmentations include horizontal image flipping, random cropping, and HSV space modulation. I took us 30 minutes to train the object detector using only real data, and around 24 hours using generated data.

\subsection{Results and discussion}

The performance of the proposed approach was tested on a test subset. As described in Section \ref{experiment} we test the quality of generated images by training an object detector on real and generated images. We used an average precision (AP) metric for assessing the performance of an object detector on a test set. The object threshold of YOLO was 0.001 while the non-maximum suppression threshold was 0.5 and the intersection over union was 0.5. Generated images used in this test were not handpicked but randomly generated.

Detailed results can be seen in Table \ref{table:results}. Using the C/P method for image synthesis did not provide any improvements in the detection. When training only on C/P images we acquire a very bad result of only 47.42\% AP on the same test set. When training on the combination of both real and C/P images we again do not get any improvements. The reason could be that this data has some artifacts when compared to the real images. Although visually, both images generated with GAN and with the copy/paste method look realistic, the object detector tends to learn wrong features and can not converge to a better model than the one trained on real images. However, we achieved an improvement with GAN generated images when opposed to training the object detector with only real images. An improvement of 1\% has been achieved while training only on GAN generated images, and an improvement of almost 5\% of AP was achieved when training on a combined dataset of real and images generated with our GAN. As a reference, experiments with two versions of the GAN without the object detector discriminator and without position mask and image concatenation in the discriminator were made. Both versions perform worse than our DetectionGAN.

\begin{table}[width=.9\linewidth,cols=6,pos=ht]
\caption{Results of training the object detector on different training datasets}\label{table:results}
\begin{tabular*}{\tblwidth}{@{} LCC@{} }
\toprule
TRAINING DATA &
AP (\%) & 
TRAINING DATA SIZE \\
\midrule
\midrule
Real & 70.96 & 2,278 \\
\midrule
C/P & 47.42 & 200,000  \\
C/P + real & 69.93 & 202,278 \\
\midrule
GAN w/o conc. & 39.90 & 200,000 \\
GAN w/o conc. + real & 69.92 & 202,278 \\
GAN w/o yolo & 62.22 & 200,000 \\
GAN w/o yolo + real & 69.3 & 202,278 \\
\midrule
DetectionGAN & 72.11 & 200,000 \\
DetectionGAN + real & \textbf{75.65} & 202,278 \\
\bottomrule
\end{tabular*}
\end{table}

This experiment proves that it is important to have the most realistic data as it is possible to achieve an improvement. It illustrates the complexity of the problem of generating synthetic data for training the object detector. Our proposed GAN is able to generate highly realistic data that can help improve the object detector.

\section{Conclusion and future work}
\label{conclusion}

In this paper, we propose a novel generative adversarial network for generating highly realistic B-scans (images) from position mask images. Our GAN generates highly realistic ultrasonic images from position masks that can be used to train an object detector. We achieved an improvement of almost 5\% while training on a combination of generated and real data. We also developed a copy/pasting method for data augmentation that proved the complexity of improving the average precision of an object detector with additional data. As we didn't cherry-pick GAN generated images all of the generated images were proven to be of high quality. 

With the increasing problem of lack of data and advances in generating high-quality synthetic data, networks such as our GAN could be used in many science and industry fields. In future work, our GAN should be tested using different object detectors.

\section{Acknowledgments}
This research was co-funded by the European Union through the
European Regional Development Fund, under the grant
KK.01.2.1.01.0151 (Smart UTX).

\printcredits

\bibliographystyle{elsarticle-num}

\bibliography{./bibtex/bib/bibliography.bib}


\bio{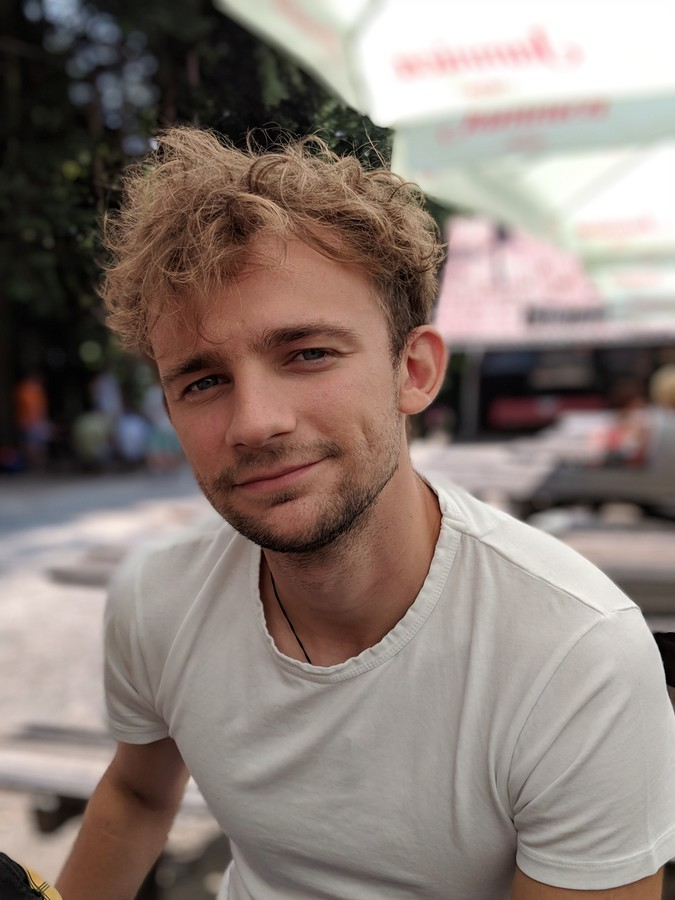}
\textbf{Luka Posilović} received his M.Sc. from the University of Zagreb, Faculty of Electrical Engineering and Computing in 2019. He is currently working as a young researcher in an Image Processing Group in the Department of Electronic Systems and Information Processing and working on his Ph.D. on the same University. His research interests include visual quality control, deep learning object detection and synthetic image generation.
\endbio

\bio{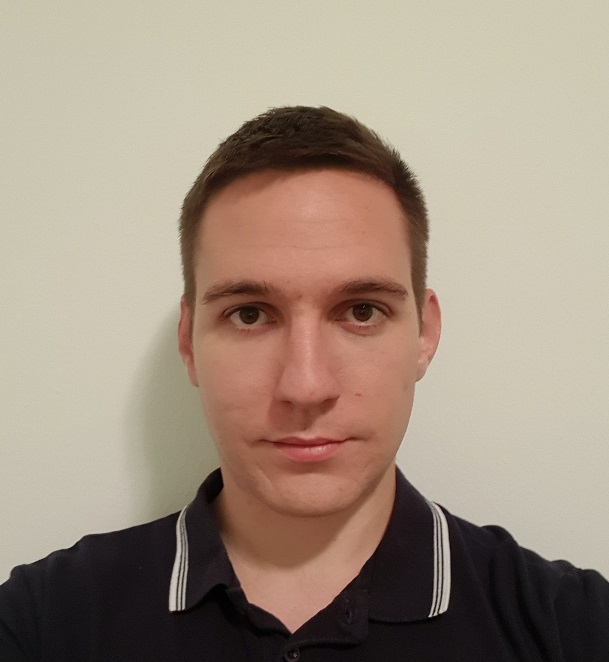}
\textbf{Duje Medak} received his M.Sc. from the University of Zagreb, Faculty of Electrical Engineering and Computing in 2019.
He is currently pursuing the Ph.D. degree in the same faculty while working as a researcher in the Image Processing Group
in the Department of Electronic Systems and Information Processing. His research interests include image processing,
image analysis, machine learning and deep learning. His current research interest includes deep learning object detection
methods and their application in the non-destructive testing (NDT) domain.
\endbio

\bio{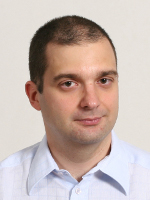}
\textbf{Marko Subašić} is an associate professor at the Department for
Electronic Systems and Information Processing,
Faculty of Electrical Engineering and Computing,
University of Zagreb, and has been working there from 1999. He received his Ph.D. degree
from the Faculty of Electrical Engineering and
Computing, University of Zagreb, in 2007. His field of research is image processing and analysis and neural networks with a particular interest in image segmentation, detection techniques, and deep learning
\endbio

\bio{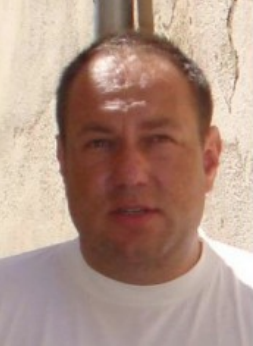}
\textbf{Marko Budimir} received his M.Sc. of physics at University of Zagreb, Faculty of Science in 2000., and his Ph.D. at Ecole Polytechnique Federale de Lausanne in Switzerland in 2006. He worked at EPFL from 2006. till 2008. From 2008. he is working at the Institute of Nuclear Technology (INETEC). He coordinated many key projects at INETEC and although he is a key person in a company of industry sector he is still working close to the field of science.
\endbio

\bio{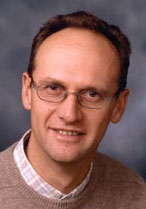}
\textbf{Sven Lončarić} received his M.Sc. in 1989. He got a Fulbright scholarship and a Ph.D. in 1994. in the University of Cincinnati, USA in the field of image processing and analysis. He is a full professor of electrical engineering and computer science with permanent tenure at Faculty of Electrical Engineering and Computing, University of Zagreb, Croatia. His areas of research interest are digital image processing and computer vision. He was principal investigator on a number of research and development projects in image processing and analysis including applications in medicine, automotive industry, and visual quality inspection.
\endbio

\end{document}